\documentclass{PoS}
\newcommand{\beq}{\begin{equation}}
\newcommand{\eeq}{\end{equation}}
\newcommand{\cA}{{\cal A}}
\newcommand{\cAb}{{\overline{\cal A}}}
\newcommand{\cF}{{\cal F}}
\newcommand{\cFb}{{\overline{\cal F}}}
\newcommand{\cD}{{\cal D}}

\newcommand{\cDb}{{\overline{\cal D}}}
\newcommand{\phib}{{\overline{\phi}}}
\newcommand{\etab}{{\overline{\eta}}}
\newcommand{\psib}{{\overline{\psi}}}
\newcommand{\cU}{{\cal U}}
\newcommand{\cUb}{{\overline{\cal U}}} 
\newcommand{\KD}{{K\"{a}hler-Dirac }}
\newcommand{\cN}{{\cal N}}
\newcommand{\Tr}{{\rm Tr\;}}

\title{Twisted lattice supersymmetry and applications to AdS/CFT}

\ShortTitle{Twisted lattice supersymmetry}

\author{\speaker{Simon Catterall}\thanks{Supported in part by DOE grant DE-FG02-85ER40237}\\
        Department of Physics, Syracuse University, Syracuse NY 13244\\
        E-mail: \email{smcatterall@gmail.com}}

\abstract{I review recent approaches to constructing supersymmetric lattice theories focusing
in particular on the concept of topological twisting. The latter technique is shown to expose
a nilpotent, scalar supersymmetry which can be implemented exactly in the lattice theory.
Using these ideas a lattice action for $\cN=4$ super Yang-Mills in four dimensions
can be written down which
is gauge invariant, free of fermion doublers and respects one out of a total of 16 continuum
supersymmetries. It is shown how these exact symmetries together with the large point
group symmetry of the lattice strongly constrain the possible counterterms needed to renormalize
the theory and hence determine how much residual fine tuning will be needed to
restore all supersymmetries in the continuum limit. We report on progress to
study these renormalization effects at one loop. We go on to give
examples of applications of these supersymmetric lattice theories to explore the
connections between gauge theories and gravity.}

\FullConference{The XXVIII International Symposium on Lattice Field Theory, Lattice2010\\
		June 14-19, 2010\\
		Villasimius, Italy}

\begin{document}

\section{Introduction}
In recent years there has been a resurgence of interest in the problem of
formulating supersymmetric
lattice theories. This has largely been driven by
the realization that in certain classes of theory a discretization
can be employed which preserves a subset of
the continuum supersymmetry.

In this talk we will review some of these basic features concentrating on
arguably the most interesting example; namely $\cN=4$ super Yang-Mills in four dimensions.
This theory, in addition to being one of the few finite four dimensional field theories, 
plays a crucial role in the original
AdS/CFT correspondence which posits an exact equivalence between the gauge theory
and a {\it dual} gravitational theory in five dimensional AdS space. The ability
both to define and ultimately
study
this theory nonperturbatively is clearly very exciting since it allows us
to test a variety of {\it holographic} conjectures connecting this
theory and its dimensional reductions to a set of different gravitational theories.
If one is optimistic it is possible that the existence of a non-perturbative construction of
this theory may lead to new insights into the nature of
quantum gravity itself.

The lattice theories I will describe have been derived using two broadly different
strategies; by careful discretization of a topologically twisted form of
the target continuum theory and also via a carefully chosen orbifolding
procedure applied to a zero dimensional mother 
theory obtained by dimensional reduction of the same target theory. Remarkably
these two approaches have been
shown to lead to essentially the same lattice theories which leads one
to suspect that the lattice constructions may be somewhat unique. Many people
have contributed to one or both of these approaches; indeed there have been more than
150 papers published on the subject in the last decade. Because of this
I have not tried to be comprehensive in citing most of the literature on
exact lattice supersymmetry but merely refer the
interested reader to the recent reviews \cite{Catterall:2009it,Giedt:2009yd} for
a more comprehensive bibliography.

In addition a body of recent work has attempted to study theories which cannot
be treated using the methods described here -- for studies of $\cN=1$ super
Yang-Mills using both Wilson and domain wall fermions see \cite{Giedt:2008xm,Endres:2009yp,Demmouche:2010sf} and
for
treatments of the Wess Zumino model in four dimensions see \cite{Bonini:2004pm, Feo:2005ex,Chen:2010uca}. There have also
been efforts directed at constructing twisted models which preserve all continuum supersymmetries
\cite{D'Adda:2010pg, D'Adda:2005zk,D'Adda:2007ax}. I will not talk about any of these other
interesting topics here.

I will start with a few comments concerning the traditional difficulties of
formulating a lattice theory with exact supersymmetry,
explain the concept of twisting and how it allows us to
circumvent most of the usual difficulties and then I will
describe the most important elements in the
construction of the $\cN=4$ lattice theory. 
I will show how the exact symmetries of the lattice theory powerfully constrain
the form of the renormalized action and hence the nature of the
continuum limit. I will go on to describe the elements of
an ongoing program to study the weak coupling structure of the theory using lattice
perturbation theory. This work is a collaboration with 
Eric Dzienkowski, Joel Giedt, Anosh Joseph and Robert
Wells \cite{us}.

Finally we will turn to applications of this lattice theory
which hinge on exploring the possible holographic connections between strongly
coupled supersymmetric
gauge theories and (super)gravity. As you will see, these initial studies have shown that practical calculations,
using the familiar algorithms used for lattice QCD, can indeed 
be used successfully to study the non-perturbative
regime of these supersymmetric theories.

\section{Why lattice supersymmetry is hard}

The problem of formulating supersymmetric theories on lattices has a long
history going back to the earliest days of lattice gauge theory. However,
after initial efforts failed to produce useful supersymmetric lattice
actions the topic languished for many years. Indeed a folklore
developed that supersymmetry and the lattice were
mutually incompatible. However, the application of new ideas and tools
drawn from other areas of theoretical physics has shown this folklore to
be incorrect and as this talk will show
for certain classes of theory the problem has been solved.\footnote{Solved in the
sense that a supersymmetric lattice theory exists which targets a given
continuum supersymmetric theory. Whether the lattice theory flows to this
continuum theory in the continuum limit without additional fine tuning is still
an open problem in certain cases including as we will see $\cN=4$ YM}

First, let me explain why discretization of supersymmetric theories
resisted solution for so long. The central problem is
that naive discretizations of continuum supersymmetric theories break
supersymmetry completely and radiative effects lead to a profusion of
{\it relevant} supersymmetry breaking counterterms in the
renormalized lattice action. The coefficients to these counterterms must
then be carefully fine tuned as the lattice spacing is sent to zero
in order to arrive at a supersymmetric theory in the continuum
limit.
In most cases this is both unnatural and practically impossible -- particularly
if the theory contains scalar fields. 

Of course, one might have
expected problems -- the supersymmetry algebra is an extension of
the Poincar\'{e} algebra which is explicitly broken on the lattice. 
Specifically,
there are no infinitesimal translation generators on a discrete spacetime
so that the algebra $\{Q,\overline{Q}\}=\gamma_\mu p_\mu$ is already
broken at the classical level.
Equivalently it is a straightforward exercise to show that a naive
supersymmetry variation of a naively discretized supersymmetric theory
fails to yield zero as a consequence of the failure of the Leibniz rule
when applied to lattice difference operators. The basic idea of the
new approaches is to maintain 
a particular subalgebra of the full supersymmetry algebra
in the lattice theory. The hope is that this exact symmetry will
constrain the effective lattice action
and protect the theory from 
dangerous susy violating counterterms.

\section{Topological twisting}

The simplest way to expose a subalgebra which is
compatible with discretization is to consider not the original
target theory but a so-called {\it twisted} variant of it. The basic idea of twisting goes back
to Witten in his seminal paper on topological field theory \cite{Witten:1988ze}
but actually had been anticipated in earlier work on
staggered fermions \cite{Elitzur:1982vh}. Indeed at the free level the final
lattice fermion action we will construct can be mapped exactly into that of (reduced) staggered fermions.
 
In our context the idea of twisting is to decompose the fields of the theory
in terms of representations not of the original (Euclidean)
rotational symmetry $SO_{\rm rot}(D)$ but a 
twisted rotational symmetry which is the
diagonal subgroup of this symmetry and an $SO_{\rm R}(D)$ subgroup
of the R-symmetry of the theory.
\beq
SO(D)^\prime={\rm diag}(SO_{\rm Lorentz}(D)\times SO_{\rm R}(D))
\eeq
To be explicit consider the case where the total number of
supersymmetries is $Q=2^D$.
In this case I can treat the supercharges of the twisted theory
as a $2^{D/2}\times 2^{D/2}$ matrix $q$. This matrix can
be expanded on products of gamma matrices
\beq
q=Q I+Q_\mu \gamma_\mu+Q_{\mu\nu}\gamma_\mu\gamma_nu+\ldots\eeq
The $2^D$ antisymmetric tensor components that arise in
this basis are the twisted
supercharges and satisfy a corresponding
supersymmetry algebra following from the original algebra
\begin{eqnarray}
Q^2&=&0\\
\{Q,Q_\mu\}&=&p_\mu\\
&\cdots&
\end{eqnarray}
The presence of the nilpotent scalar supercharge $Q$ is most important;
it is the algebra of this charge that we can hope to translate to
the lattice. The second piece of the algebra expresses the fact that
the momentum is the $Q$-variation of something which makes plausible
the statement that the energy-momentum tensor and hence the entire
action can be written in $Q$-exact form\footnote{Actually in the case
of $\cN=4$ there is an additional $Q$-closed term needed}.
Notice that an action written in such a $Q$-exact form is trivially
invariant under the scalar supersymmetry provided the latter remains
nilpotent under discretization.

The rewriting of the supercharges in terms of twisted variables can
be repeated for the fermions of the theory and yields a set
of antisymmetric tensors $(\eta,\psi_\mu,\chi_{\mu\nu},\ldots)$
which for the case of $Q=2^D$ 
matches the number of components of a real \KD field.
This repackaging of the fermions of the theory into a \KD field
is at the heart of how the discrete theory avoids fermion doubling
as was shown by Becher, Joos and Rabin in the early days of
lattice gauge theory \cite{Rabin:1981qj,Becher:1982ud}. 

It is important to recognize
that the transformation to twisted variables corresponds to a simple
change of variables in flat space -- one more suitable to
discretization. A true topological field theory only results when
the scalar charge is treated as a true BRST charge and attention is
restricted to states annihilated by this charge. In the language of
the supersymmetric parent theory such a restriction corresponds to a projection
to the vacua of the theory. It is {\it not} employed in these
lattice constructions.

\section{Twisted $\cN=4$ super Yang-Mills}

This theory satisfies our requirements for supersymmetric latticization;
its R-symmetry possesses an $SO(4)$ subgroup corresponding to rotations
of its four degenerate Majorana fermions into each other which may
be twisted with the Euclidean Lorentz symmetry. The twisted fermions being
initially spinors under both the R symmetry and the Lorentz group transform as
integer spin antisymmetric tensors after twisting. In a similar fashion, 4 of the
scalars which formed a vector representation of the R symmetry
transform as vectors under the twisted rotation group. Indeed, as we will see they
can be packaged with the gauge fields as part of a complexified connection.

The theory can be written in twisted form as
\beq S=\frac{1}{g^2}\left(S_{\rm exact}+S_{\rm closed}\right)\eeq
where
\beq
S_{\rm exact}=Q
\int\Tr\left(\chi_{ab}\cF_{ab}+\eta [ \cDb_a,\cD_a ]-\frac{1}{2}\eta
d\right)\label{4daction}\eeq
In this expression the indices run not over $1\ldots 4$ as one would expect but
$1\ldots 5$ corresponding to a theory with a twisted $SO(5)$ invariance. Indeed the compact expression
given in eqn.~\ref{4daction} is most easily understood as a naive
dimensional reduction of a {\it five}
dimensional theory. The ten bosons of $\cN=4$ YM are thus mapped after
dimensional reduction to a complex scalar $\cA_5$ and a complex gauge field $\cA_\mu,\mu=1\ldots 4$ in which
the remaining four scalars of $\cN=4$ appear as the imaginary parts of the gauge field. This five component notation will
prove particularly useful later when we discretize the theory on a lattice.
The 16 fermions
$(\eta,\psi_a,\chi_{ab})$ naturally decompose on the fields required for a
4d \KD field. Explicitly
\begin{eqnarray}
\cA_a&\to&\cA_\mu\oplus\phi\nonumber\\
\cF_{ab}&\to& \cF_{\mu\nu}\oplus\cD_\mu \phi\nonumber\\
\left[\cDb_a,\cD_a\right]&\to&\left[\cDb_\mu,\cD_\mu\right]\oplus
\left[\phib,\phi\right]\nonumber\\
\psi_a&\to&\psi_\mu\oplus\etab\nonumber\\
\chi_{ab}&\to&\chi_{\mu\nu}\oplus\psib_{\mu}
\end{eqnarray}
where we will employ the convention that Greek indices run from one to
four and are reserved for
four dimensional tensors while Roman indices refer to the original
five dimensional theory. The resulting action is just the one constructed by
Marcus \cite{Marcus} and later referred to as the GL-twist \cite{Kapustin:2006pk}.
This twisted action is well known to be fully equivalent to the usual
form of ${\cal N}=4$ in flat space.

The nilpotent transformations associated
with $Q$ are given explicitly by
\begin{eqnarray*}
Q\; \cA_a&=&\psi_a\nonumber\\
Q\; \psi_a&=&0\nonumber\\
Q\; \cAb_a&=&0\nonumber\\
Q\; \chi_{ab}&=&-\cFb_{ab}\nonumber\\
Q\; \eta&=&d\nonumber\\
Q\; d&=&0
\end{eqnarray*}
The second component of the action $S_{\rm closed}$ takes the form
\beq
S_{\rm closed}=-\frac{1}{2}\int\epsilon_{abcde}\chi_{ab}\cDb_c\chi_{de}\label{Qclosed}\eeq
and is supersymmetric on account of the (five dimensional) Bianchi identity.

Performing the $Q$-variation and integrating out the auxiliary field
$d$ yields
\beq
S=\frac{1}{g^2}\int\Tr\left(-\cFb_{ab}\cF_{ab}+\frac{1}{2}[ \cDb_a, \cD_a]^2-
\chi_{ab}\cD_{\left[a\right.}\psi_{\left.b\right]}-
\eta \cDb_a\psi_a\right)-\frac{1}{2g^2}\int\epsilon_{abcde}\chi_{ab}\cDb_c\chi_{de}\eeq

\section{Discretization}

The prescription for discretization is somewhat natural. Complex
gauge fields are represented as complexified Wilson gauge links
$\cU_a(x)=e^\cA_\mu(x)$ living on links of a lattice.
The most natural lattice possesses {\it five}
equivalent basis vectors satisfying $\sum_{i=1}^5 e_i=0$
and is called $A_4^*$. It has a the large
point group symmetry $S^5$ - far larger than the usual hypercubic group.
This will prove important later
when we consider the renormalization of the
theory.

The link fields
transform in the usual way under $U(N)$ lattice gauge transformations
\beq
\cU_a(x)\to G(x)\cU_a(x)G^\dagger(x+a)\eeq
Supersymmetric invariance
then implies that $\psi_a(x)$ live on the same links
and transform identically. 
The scalar fermion $\eta(x)$ is clearly most naturally associated with
a site and transforms accordingly
\beq \eta(x)\to G(x)\eta(x)G^\dagger(x)\eeq
The field $\chi_{ab}$ is slightly more difficult. Naturally as a 2-form
it should be associated with a plaquette. In practice we introduce additional
links running from $x\to x+a+b$ and let $\chi_{ab}$
lie {\it with opposite orientation} along those links. This
choice of orientation will be necessary to ensure gauge
invariance.

To complete the discretization we need to describe how continuum derivatives
are to be replaced by difference operators. A natural technology for
accomplishing this in the case of adjoint fields was developed many
years ago and yields expressions for
the derivative operator applied to arbitrary
lattice p-forms \cite{Aratyn}. 
In the case discussed here we need just three derivatives given by
the expressions
\begin{eqnarray}
\cD^{(+)}_a f_a&=&
\cU_a(x)f_b(x+a)-f_b(x)\cU_a(x+b)\\
\cDb^{(-)}_a f_a&=&f_a(x)\cUb_a(x)-\cUb_a(x-a)f_a(x-a)\\
\cDb^{(-)}_c f_{ab}&=&f_{ab}(x)\cUb_c(x-c)-
\cUb_c(x+a+b-c)f_{ab}(x-c)
\label{derivs}
\end{eqnarray} 
The lattice field strength is given by 
the gauged forward difference $\cF_{ab}=\cD^{(+)}_a \cU_b$
and is automatically antisymmetric in its indices.
Furthermore it transforms like
a lattice 2-form and
yields a gauge invariant loop on the lattice when contracted
with $\chi_{ab}$.
Similarly the covariant backward difference appearing in $\cDb_a \cU_a$
transforms as a 0-form or site field and hence can be contracted with
the site field $\eta$. The expression in the third line of eqn.~\ref{derivs} is carefully
chosen so that the $Q$-closed term is gauge invariant on the lattice -- the epsilon tensor
forces all the indices to be distinct in eqn.~\ref{Qclosed} and since the
basis vectors sum to zero the expression can be
seen to correspond yet again to a closed loop. Furthermore the lattice field strength satisfies
a Bianchi identity just as for the continuum so the term is also $Q$-supersymmetric.

This use of forward and backward difference operators guarantees that the
solutions of the theory map one-to-one with the solutions of the continuum
theory and hence fermion doubling problems are evaded \cite{Rabin:1981qj}.
Indeed, by introducing a lattice with half the lattice spacing one can
map this \KD fermion action into the action for staggered fermions. 
Notice that, unlike the case of QCD, there is no rooting problem in
this supersymmetric construction since the additional fermion
degeneracy is already required by the continuum theory.

To summarize; the lattice theory we have constructed by
discretization of a twisted reformulation of the target
SYM theory is gauge invariant, $Q$-supersymmetric and 
free of fermion doubling problems. However, many questions remain to be answered; is the theory rotationally
invariant in the continuum limit, does it flow to the target theory with full supersymmetry as the
lattice spacing is reduced and does the classical moduli space
survive under quantum correction. We now summarize what is known about these issues by examining the
theory using perturbative methods.

\section{Renormalization of twisted lattice theory}

Before we embark on a general perturbative analysis of this lattice theory it is instructive to try to 
ascertain what kinds of counter terms are permitted by the lattice symmetries. In the case of $A_4^{*}$ lattice, these symmetries are exact $Q$ supersymmetry, gauge invariance and the $S^5$ point symmetry
of the $A_4^*$ lattice.
In fact,
the $U(N)$ lattice gauge theory also has a second fermionic symmetry,  given by
\beq
\eta \rightarrow \eta + \epsilon {\mathbb I}_N, \qquad  \delta (\rm all \; other \; fields) =0 
\eeq
where $\epsilon$ is an infinitesimal Grassmann parameter. This acts as a further constraint on
the structure of the renormalized lattice action. 

In practice we are primarily interested in  relevant or marginal  operators; that is operators whose mass dimension is less than or equal to four. We will see that the set of relevant counterterms in the lattice theory is rather short -- the lattice symmetries, gauge invariance in particular, being extremely restrictive in comparison to the equivalent situation in the continuum. The argument starts by assigning canonical dimensions to the fields 
$[\cU_a]=[\cUb_a]=1$, $[\Psi]=\frac{3}{2}$ and $[Q]=\frac{1}{2}$ where $\Psi$ stands for any of the twisted fermion fields $(\lambda,\psi^m,\xi_{mn})$. Invariance under $Q$ restricts the possible counterterms to be either of a  $Q$-exact form, or  of  $Q$-closed form. There is only one $Q$-closed operator permitted by the lattice symmetries and it is already present in our bare lattice action. A possible renormalization of this fermion kinetic term is hence allowed. Beyond that the exact lattice supersymmetry forces us to look at the set of $Q$-exact counterterms.    

Any such counterterm must be of the form $\mathcal{O}=Q\Tr(\Psi f(\cU,\cUb))$. There are thus no terms permitted by symmetries with dimension less than two. In addition gauge invariance tells us that each term must correspond to the trace of a closed loop on the lattice. The smallest dimension gauge invariant operator is then just $Q(\Tr\psi_a\cUb_a)$. But this vanishes identically since both $\cUb_a$ and $\psi_a$ are singlets under $Q$. No dimension $\frac{7}{2}$ operators can be constructed with this structure and we are left with just dimension four counterterms. Notice, in particular that lattice symmetries permit no simple fermion bilinear mass terms.  However, gauge invariant fermion bi-linears with link field insertions are possible and their effect should be accounted for carefully.  
Possible  dimension four operators are, schematically, 
\begin{eqnarray}
L_1&=& Q \Tr (\chi_{ab}\cU_a \cU_b)  \cr
L_2&=&  Q \Tr (\eta \cDb_a \cU_a )\cr 
L_3&=&  Q \Tr (\eta \cU_a \cUb_a)\cr
L_4&=&  Q \Tr (\eta)\Tr(\cU_a \cUb_a)
\label{ops}
\end{eqnarray}
The  first operator  can be simplified on account of the antisymmetry of $\chi_{ab}$ to simply $Q(\chi_{ab}\cF_{ab})$, which again is nothing but 
one of the continuum $Q$-exact terms present in the bare action. The second term  in (\ref{ops}) also corresponds to one of the $Q$-exact terms in the bare action. However the third term $L_3$  is a new operator not present in the bare Lagrangian and the same is true for the final double-trace operator $L_4$. Both of these operators transform non-trivially under the fermionic shift symmetry, but a linear combination  of the two 
\beq 
D =  L_3 - \frac{1}{N} L_4  
\label{ren4}
\eeq
is invariant under the shift symmetry with $N$ the  number of colors of the gauge group  $U(N)$. 
 
By these arguments it appears that the only relevant counterterms correspond to renormalizations of 
marginal operators already present in the bare action together with $D$. The effective lattice action taking the form
\beq S=\sum_x Q\left(\alpha_1 \chi_{ab}F_{ab}+\alpha_2 \psi_a\cDb_a \eta+\frac{\alpha_3}{2}\eta d\right) +\alpha_4 S_{\rm closed}+\alpha_5 D\eeq
This implies that a maximum of
four couplings might require log tuning to achieve a continuum limit invariant under
full supersymmetry with only the dangerous mass term potentially needing
power law tuning with the lattice spacing. In fact we will now argue that the the 
dangerous mass terms in $D$ are in fact absent to
all orders in perturbation theory reducing any ``fine'' tuning to just four independent
logarithmic terms.
The absence of mass terms will be shown in the next section using by a general
background field calculation of the effective potential in the lattice theory. 

\section{Effective action}

Here we expand about a constant commuting background corresponding to a generic point in the
moduli space space and a solution of the classical equations of motion.
\beq
\cU_a(n) = \cU_a + i \cA_a(n), \quad \cUb_a(n) = \cUb_a - i \cAb_a(n)
\label{Urep}
\eeq
with $[\cU_a,\cUb_b]=0$.
We further add a Lorentz gauge fixing term to the action of the form 
\beq \frac{1}{4\alpha}\Tr (\cDb_a \cA_a+\cD_a\cAb_a)^2 \eeq
Choosing the gauge $\alpha=1$, we can show that the quadratic part of the
bosonic action then takes the form
\beq
S_B = -2 \sum_n \Tr \cA_b \cDb_a^- \cD_a^+ \cA_b
\eeq
Here the covariant derivatives depend only on the
constant commuting classical background fields $\cU_a, \cUb_a$.
This gauge fixing functional leads to the quadratic ghost action
\beq
S_G = \sum_n \Tr \overline{c} \cDb_a^- \cD_a^+ c
\eeq
The quadratic fermionic part of the action is given by the corresponding
expression in eqn.~\ref{4daction}, except that now the covariant
derivatives now depend only on the background fields.

Since the background is constant, we can pass to momentum
space in which the action separates into terms for each
mode $k$.  The $16 \times 16$ fermion matrix $M(k)$ (see next section for details) for the mode $k$ then can be shown
(using Maple to compute the determinant) to satisfy
\beq
\det \,M(k) = {\rm det}^8( \cDb_a^-(k) \cD_a^+(k) )
\eeq
Going back to position space we obtain
\beq
{\rm Pf}\,M = {\rm det}^4( \cDb_a^- \cD_a^+ )
\eeq
The ghosts add another factor of ${\rm det}(\cDb_a^- \cD_a^+)$ which together with the
twisted fermions
is just what is needed to cancel the factor
of ${\rm det}^5(\cDb_a^- \cD_a^+)$ which comes from the bosons
in the denominator.  In conclusion, for a constant,
commuting background, the one-loop effective action is zero - bosonic and
fermionic fluctuations cancel exactly. 
Thus the moduli space is not
lifted in this analysis and hence there can be no boson or fermion masses at one
loop. We have verified this in the next section by explicit computation of the fermion
self-energies which all vanish as the external momentum goes to zero.

Furthermore, 
we expect that we can extend this result to 
all loops since the full effective action of the lattice is related to the
partition function of the theory $Z$. The latter 
should be a topological invariant since it is just the Witten index for
the theory if periodic boundary conditions are used for all fields.
This fact implies that $Z$ that it can
be computed exactly at one loop. Indeed,
Matsuura uses similar arguments to show that the vacuum 
energy of supersymmetric orbifold theories with 
four and eight supercharges remains zero to all orders in the coupling \cite{Matsuura:2007ec}.
Thus we conclude that boson and scalar masses remain zero to all orders
in the coupling constant.

To proceed further we need to setup a perturbative calculation for the 
remaining coefficients $\alpha_i,i=1\ldots 4$.

\section{One loop lattice perturbation theory}

Three of the four coefficients $\alpha_1$, $\alpha_2$ and $\alpha_4$
can be computed by examining the fermion self energy diagrams. 
The basic ingredients required
for the perturbative computation of these quantum corrections are the propagators for the fermion and boson fields and the
vertices. These are derived in the usual way by substituting the expression for the
$\cU$ fields given in eqn.~\ref{Urep} into the lattice action and identifying the quadratic pieces with the propagators
and the vertices from the interactions.
If we adopt the gauge fixing described in the previous section only a single non-zero
boson propagator is required which takes the form in momentum space
\beq<\cAb_a^C(k)
\cA_b^D(-k)>=\frac{1}{\hat{k}^2}\delta_{ab}\delta^{CD}\eeq
The fermion propagator takes the form of a $16\times 16$ 
block matrix which acts on the twisted fields $\Psi_i=(\eta,\psi_a,\chi_{ab}),i=1\ldots 16$.
In momentum space it is
\beq
\langle \Psi_i^A(-k)\Psi_j^B(k) \rangle =
\frac{1}{2\widehat{k}^2}
M_{ij}(k)\delta_{AB}
\eeq
where the matrix $M$ is the discrete
\KD operator and $\widehat{k}^2=\sum_a \sin^2{\frac{k_a}{2}}$. Notice that the square of the
fermion propagator is nothing more than the boson propagator as a consequence of the
supersymmetric structure of the lattice action.
The vertices correspond to terms involving the fields $\psi\eta$, $\psi\chi$ and $\chi\chi$. 
It can be easily be shown that only
four Feynman diagrams are needed at one loop to find $\alpha_1,\alpha_2,\alpha_4$. An example of a
typical contribution to the one
loop renormalization of the $\chi-\chi$ propagator is shown below:

\begin{tabular}{cc}
\includegraphics[width=0.4\textwidth]{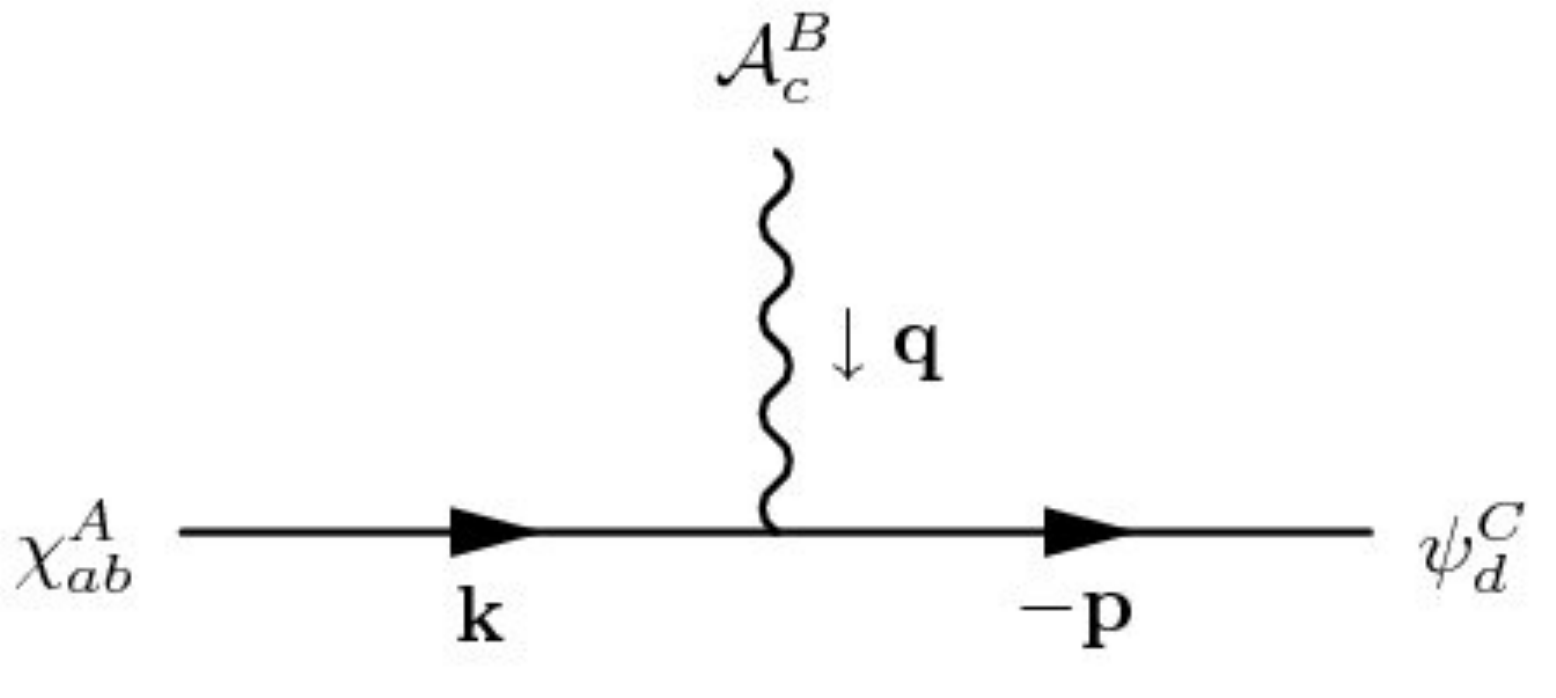}&
\includegraphics[width=0.4\textwidth]{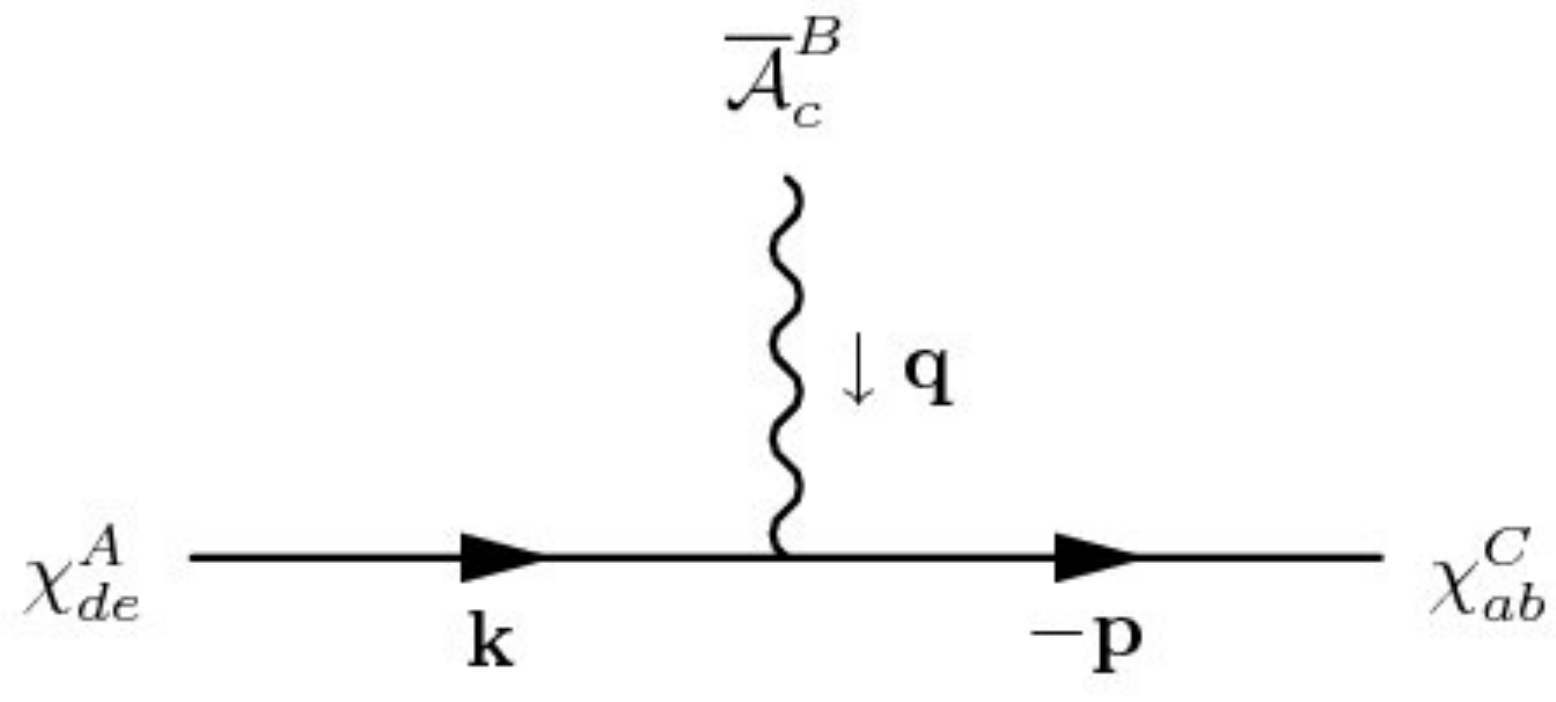}
\end{tabular}
\begin{center}\includegraphics[width=0.4\textwidth]{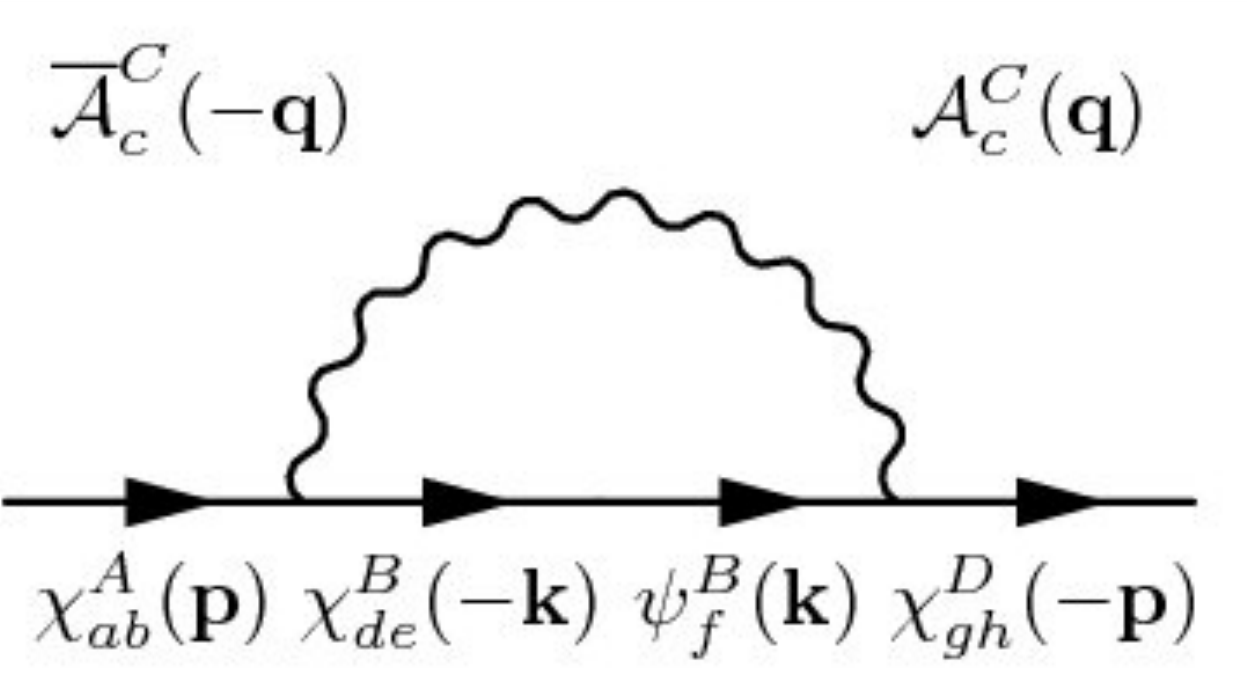}\end{center}

The corresponding expression that follows by applying the
Feynman rules to this diagram
is somewhat complicated and we refer the reader to \cite{us} for its
precise form but it can be shown to possess the following properties;
\begin{itemize}
\item It vanishes as $p\to 0$  as expected from the general arguments given in the last
section concerning the absence of mass terms.
\item It contains a logarithmic divergence $Ag^2\log{\mu a}$ which can be obtained by fitting the
infinite lattice integral for the amputated diagram as a function of the IR regular mass $\mu$.
\end{itemize}
It can shown that all the
fermion self-energy one loop terms possess the same properties. 

In renormalized perturbation theory the log divergences can then be absorbed using
wavefunction renormalization counterterms with the finite parts of the counterterms determined as
usual by choosing suitable normalization conditions on the two point functions. The coefficients of
the log divergences yield
weak coupling estimates for the anomalous dimensions of the twisted fermion fields and at the same
time
give the required tuning as $a\to 0$ of the bare parameters $\alpha_1,\alpha_2,\alpha_4$ required to 
to ensure that full supersymmetry is restored in the lattice theory at
weak coupling. A complete understanding of this issue requires in addition a computation of
the renormalization of $\alpha_3$ which can be obtained via  
the vacuum polarization of the complex gluons. 

This program sketched out in this section is near completion and we refer the interested reader to
\cite{us} for further details. We expect this one loop calculation will serve as an
important guide for numerical explorations of the theory at strong coupling.

\section{Numerical simulations}

We have conducted preliminary investigations of the theory on small lattices and confirmed that Monte Carlo
simulations of the full theory are practicable; Figure 1. illustrates that
the observed Pfaffian phase is small in the parameter
regions of interest at least for the small lattices that
have been examined so far\footnote{Similar studies have provided
evidence that the related two dimensional lattice model with 4 supercharges also does not
possess a sign problem in the continuum limit \cite{Hanada:2010qg}}\cite{Catterall:2008dv}. 
Furthermore, the flat directions of the theory corresponding to constant commuting fields
do not seem
to cause problems of convergence; the scalar fields of the
lattice theory do not appear to evolve to ever increasing values 
as the simulation progresses at least for zero temperature but remain
somewhat localized close to the origin in field space 
as can be seen in Figure 2. 
\begin{figure}
\begin{center}
\includegraphics[width=0.6\textwidth]{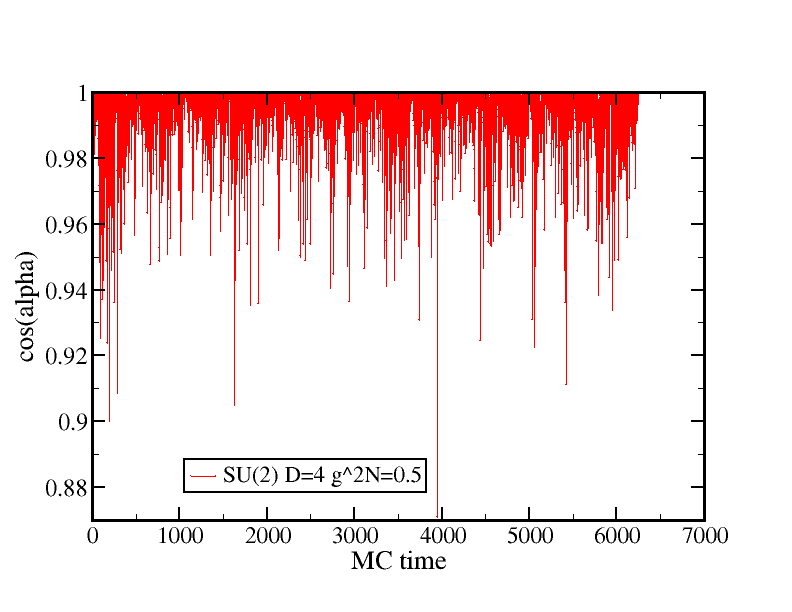}
\caption{Cosine of pfaffian phase vs Monte Carlo time}
\end{center}
\label{pfaffian}
\end{figure}

\begin{figure}
\begin{center}
\includegraphics[width=0.6\textwidth]{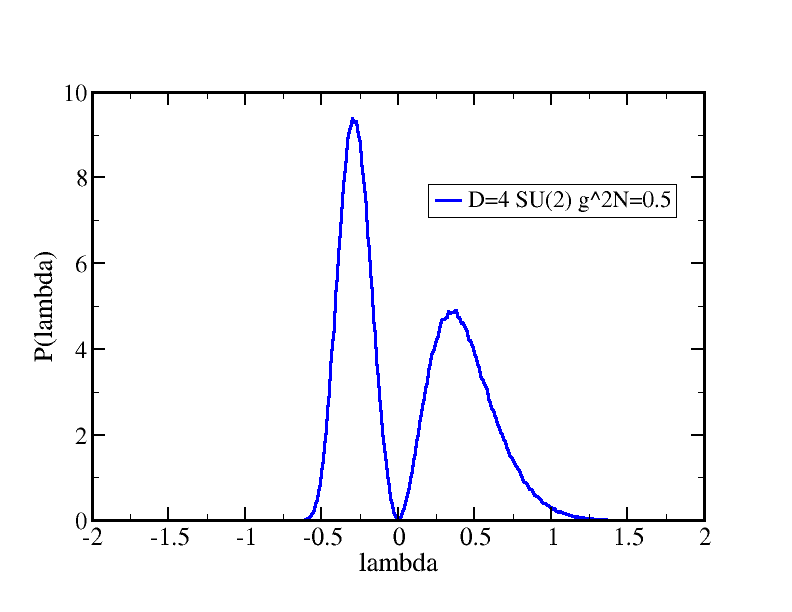}
\caption{Probability distribution of scalar eigenvalues $\lambda$ of $U^\dagger_a U_a-1$}
\end{center}
\label{flat}
\end{figure}

A parallel code has now been constructed which uses a multiple time step RHMC algorithm to handle
the integration over the twisted fermions. The code uses the MDP libraries within the
FermiQCD package to provide an interface to the MPI libraries that handle
communication between individual computation nodes.
The plot in Figure 3. below shows the performance of this code as a function of
the number of cores for a simulation of the $SU(2)$ theory on a $8^3\times 16$ lattice.

\begin{figure}
\begin{center}
\includegraphics[width=0.6\textwidth]{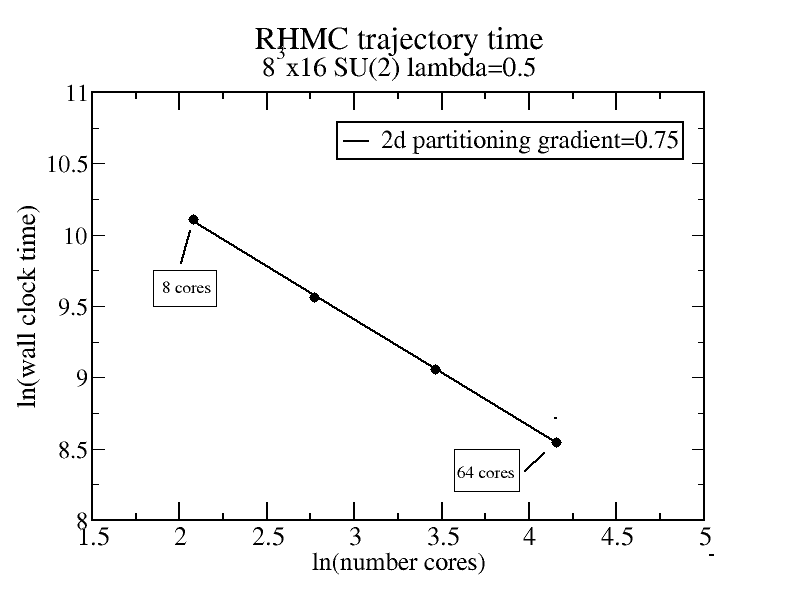}
\caption{Wall clock time for $\cN=4$ SYM model on $8^3\times 16$ vs number of cores}
\end{center}
\label{perf}
\end{figure}

This illustrates that simulations of $\cN=4$ super Yang-Mills in four dimensions using this Q-exact action are
practicable and have already started. We now turn to recent applications of this numerical work to study
conjectured dualities between dimensionally reduced versions of this theory and various supergravity
theories. These provide concrete examples of how simulations of Yang-Mills theories 
regulated using these methods can provide
insight into the properties of black holes.

\section{Applications: black holes from Yang-Mills theory}

In recent years there has been a great deal of interest in dualities between supersymmetric
Yang-Mills theories and gravitational theories. For example, the original AdSCFT correspondence has been
generalized to allow for the description of the low energy dynamics of Dp-branes in
various supergravity theories at finite temperature and the thermal behavior of $(p+1)$-dimensional
Yang-Mills theories at strong coupling and large N \cite{Itzhaki:1998dd}.

The existence of supersymmetric lattice actions for these Yang-Mills theories allows us to
test these conjectured correspondences in some detail and indeed precise computations in 
the Yang-Mills theories can in principle yield information on the stringy corrections to the
supergravity description. With this in mind in mind we have begun the study of two Yang-Mills
theories both of which are derived by dimensional reduction of the $\cN=4$ lattice theory.

The first of these is thermal Yang-Mills quantum mechanics which is thought to be dual to
a system of $D0$ branes in type iia string theory at non zero temperature. For low temperature and
large N the system should be well described by certain black hole solutions in 
iia supergravity and standard Bekenstein-Hawking arguments can be used to compute the
mean energy and entropy of these black holes. We have checked this correspondence by
simulation of the supersymmetric Yang-Mills model and find good agreement; Figure 4. shows
a plot of the mean energy in the Yang-Mills system for several values of N versus the
(dimensionless) temperature. The data points correspond to the Monte Carlo data while
the solid curve is the prediction for the corresponding black hole solution. Technical issues
related to an IR divergence of the thermal partition function make it difficult to simulate
the Yang-Mills system at very low temperatures; nevertheless the agreement is quite
impressive. Notice that the data points corresponding to the
quenched theory rapidly diverge from the black hole curve at low temperature and contrast
with the results of the dynamical fermion simulations which turn over to approach
zero at low temperature as expected for a supersymmetric system. These results are discussed
in more detail in \cite{Catterall:2008yz,Catterall:2009xn} and are in good agreement with other non lattice methods which have
also targeted this system \cite{Anagnostopoulos:2007fw,Nishimura:2009xm,Hanada:2008ez,Hanada:2008gy}.

\begin{figure}
\begin{center}
\includegraphics[width=0.6\textwidth]{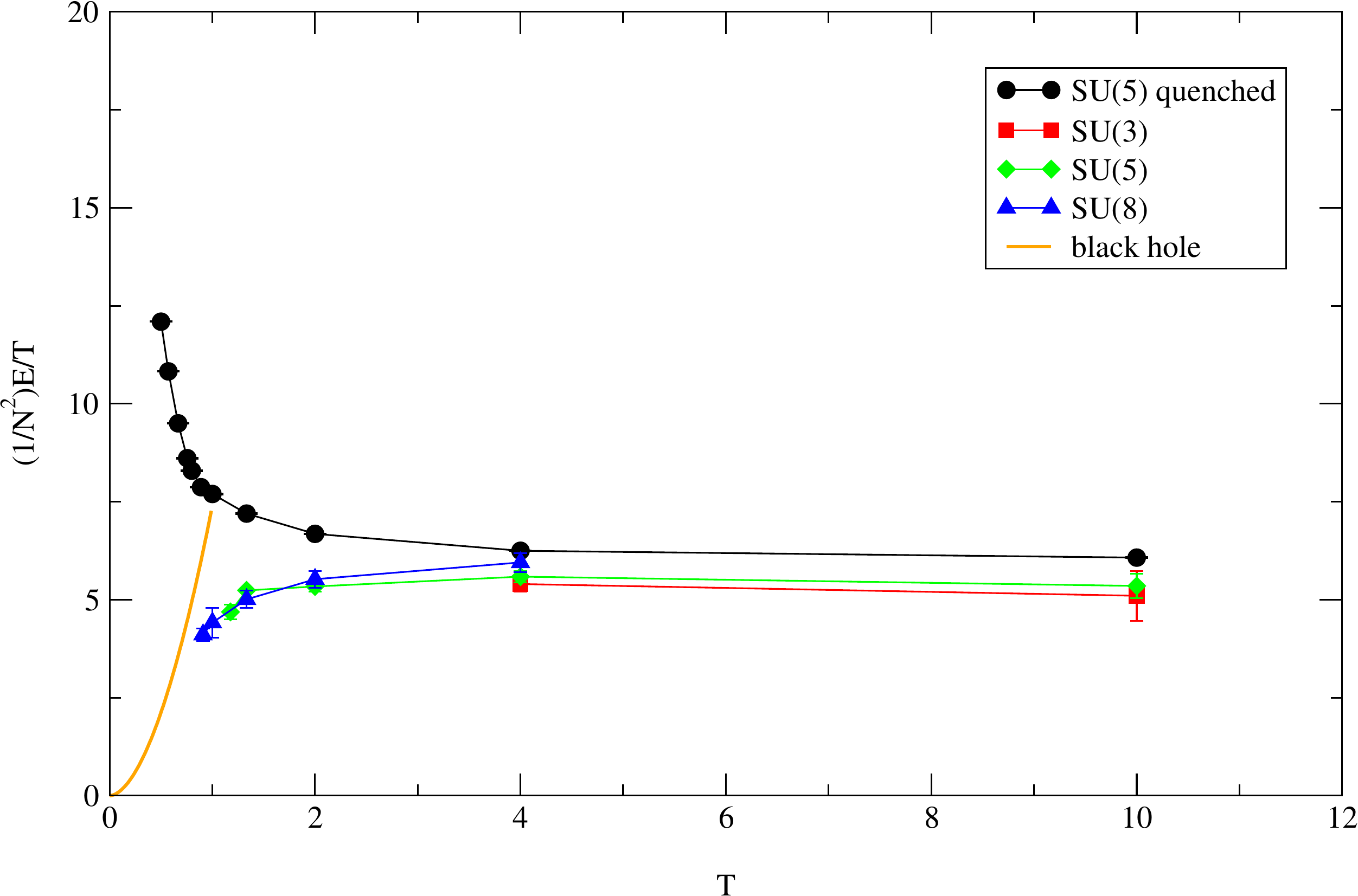}
\caption{Comparing $E_{\rm YM}$ with $E_{\rm BH}$ for $Q=16$ SYMQM}
\end{center}
\label{QM}
\end{figure}

The case of SYM in two dimensions is perhaps even more interesting; in this case the phase diagram of
the Yang-Mills theory compactified on a torus
is labeled by two parameters; the extent of the lattice in the (Euclidean)
time direction and the spatial extent. These two directions are
distinguished by the boundary conditions on the fermions; antiperiodic in the temporal
direction, periodic in the spatial direction. In the supergravity dual they correspond to
the temperature and the size of a compactified spatial direction. Two solutions of the
supergravity equations are now allowed corresponding to a spherically symmetric black
hole solution and a new type of solution with cylindical symmetry - the black string.
Which solution dominates the free energy depends on the size of the black hole horizon in
comparison to the size of the compactified dimension. If the event horizon wraps the spatial
circle the black string solution is preferred. The transition between these
different types of spacetime is referred to as the Gregory-LaFlamme instability
\cite{Gregory:1994bj}. In the dual
Yang-Mills system it can be seen as a thermal phase transition distinguished by the
spatial Polyakov line. We have investigated the phase boundary between confined and deconfined
behavior of this line as the aspect ratio of the lattice and temperature are
varied for several N \cite{Catterall:2010fx}.

Figure 5. shows a contour plot of our results in the parameter space corresponding to
the temporal $r_\tau$ and spatial extents $r_x$ rendered dimensionless by rescaling with the 't Hooft
coupling. The solid curves correspond to results obtained from the
supergravity solution in blue and the dimensionally reduced model obtained at high temperature (red).
It can be seen
that the numerically obtained phase boundary (defined as the point where $P_{\rm spatial}=0.5$)
shown by the two dashed lines for $SU(3)$ and $SU(4)$
corresponds rather well to both analytic limits and in particular, confirms the
expected parametric dependence of $r_\tau\sim c r_x^2$ given by supergravity. The coefficient $c$ which was
previously unknown from the gravity side taking the approximate value
$c\sim 3.5$.

\begin{figure}
\begin{center}
\includegraphics[width=0.5\textwidth]{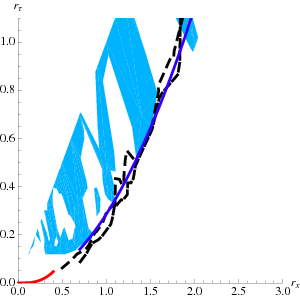}
\caption{Transition line in $P_s$ for $SU(3),SU(4)$ with fit to supergravity prediction in blue}
\end{center}
\label{string}
\end{figure}

\section{Summary}
In this talk I have surveyed some of the recent theoretical work focused on
constructing supersymmetric lattice theories and described some initial perturbative and
numerical studies of these models with particular emphasis on applications to studies of gauge gravity
duality.

Recent developments that I have not had time to cover include theoretical constructions to
couple these theories to fermions in the fundamental representation \cite{Kadoh:2009yf,Kikukawa:2008xw}, Q-exact constructions
of mass deformed YM theories \cite{Hanada:2010kt,Catterall:2010gf,Hanada:2010gs} and
connections of twisted YM to gravity \cite{Catterall:2010ng}. With these continuing theoretical developments and
the start of serious numerical studies I expect the next few years will prove to be an exciting time
for lattice supersymmetry.

\end{document}